\documentclass[12pt]{iopart}

\usepackage{amsfonts}%
\usepackage{amssymb}%
\usepackage[cal=cm]{mathalfa}
\usepackage{graphicx}
\usepackage{braket}
\usepackage{sidecap}
\usepackage{color}
\usepackage{bbm}
\usepackage{nicefrac}
\usepackage{float}
\usepackage[dvipsnames]{xcolor}
\usepackage[utf8]{inputenc}
\usepackage[normalem]{ulem}
\usepackage{nicefrac}
\usepackage{caption}
\usepackage{soul,xcolor}
\usepackage[subrefformat=parens,labelformat=simple]{subcaption}

\newcommand{\stkout}[1]{\ifmmode\text{\sout{\ensuremath{#1}}}\else\sout{#1}\fi}

\newcommand{\id}{\mathbbm{1}}

\usepackage[colorlinks=true, citecolor=blue, linkcolor=blue, urlcolor=blue]{hyperref}

\begin{document}

\setstcolor{red}

\title[Test of an entropic measurement uncertainty relation for arbitrary qubit observables]{Experimental test of an entropic measurement uncertainty relation  for arbitrary qubit observables}

\author{B\"ulent Demirel$^1$, Stephan Sponar$^1$, Alastair A. Abbott$^2$, \\Cyril Branciard$^2$, and Yuji Hasegawa$^{1,3}$}

\address{$^1$ Atominstitut, TU Wien, Stadionallee 2, 1020 Vienna, \\Austria}
\address{$^2$ University Grenoble Alpes, CNRS, Grenoble INP, Institut N\'eel, 38000 Grenoble, France}
\address{$^3$ Department of Applied Physics, Hokkaido University, Kita-ku, Sapporo 060-8628, Japan}

\ead{hasegawa@ati.ac.at}

\vspace{2pc}


\vspace{10pt}

\begin{abstract}
The uncertainty principle is an important tenet and active field of research in quantum physics. Information-theoretic uncertainty relations, formulated using entropies, provide one approach to quantifying the extent to which two non-commuting observables can be jointly measured. Recent theoretical analysis predicts that general quantum measurements are necessary to saturate some such uncertainty relations and thereby overcome certain limitations of projective measurements. Here, we experimentally test a tight information-theoretic measurement uncertainty relation with neutron spin-$\nicefrac{1}{2}$ qubits. The noise associated to the measurement of an observable is defined via conditional Shannon entropies and a tradeoff relation between the noises for two arbitrary spin observables is demonstrated. The optimal bound of this tradeoff is experimentally obtained for various non-commuting spin observables. For some of these observables this lower bound can be reached with projective measurements, but we observe that, in other cases, the tradeoff is only saturated by general quantum measurements (i.e., positive-operator valued measures) as predicted theoretically. These results showcase experimentally the advantage obtainable by general quantum measurements over projective measurements when probing certain uncertainty relations.
\end{abstract}

\section{Introduction}

The uncertainty principle was one of the first quantum phenomena discovered without any classical analogue. 
In 1927 Heisenberg presented his $\gamma$-ray microscope \textit{Gedankenexperiment}~\cite{Heisenberg1927} demonstrating that the position and momentum of an electron cannot be determined simultaneously with arbitrary precision. The famous uncertainty relation $\Delta (Q) \Delta (P)\ge \frac{\hbar}{2}$ for position $Q$ and momentum $P$~\cite{Kennard1927}, however, quantifies the accuracy with which a state can be \emph{prepared} with respect to the observables of interest, rather than the ability to jointly \emph{measure} them. For several decades, research on the uncertainty principle focused on such so-called \emph{preparation uncertainty relations}.

The advent of information theory provided novel approaches to quantifying uncertainty, such as the Shannon entropy~\cite{Shannon1948}, with wide-ranging applications~\cite{Cover06}; consequently, entropic uncertainty relations were formulated soon thereafter~\cite{Hirschman1957,Beckner1975,BialynickiBirula1975}. For finite dimensional systems, novel entropic relations such as Deutsch's~\cite{Deutsch1983} and Maassen and Uffink's inequalities~\cite{MaassenUffink1988} presented advantages, such as state-independence, over Robertson's relation $\Delta (A) \Delta (B)\ge\left|\frac{1}{2 i} \braket{\psi|[A,B]|\psi}\right|$, for arbitrary observables $A$ and $B$ and any state $\ket{\psi}$~\cite{Robertson1929}. Entropic uncertainty relations have subsequently proven useful in quantum cryptography~\cite{Koashi2006,Broadbent2016}, entanglement witnessing~\cite{Berta2010}, complementarity~\cite{Coles2014} and other topics in quantum information theory~\cite{NielsenChuang}, where entropy is a natural quantity of interest.

In recent years \textit{measurement uncertainty relations}, in the spirit of Heisenberg's original proposal, have received renewed attention. Such uncertainty relations can be subdivided into two classes: \emph{noise-disturbance relations}, which quantify the idea that the more accurately a measurement determines the value of an observable, the more it disturbs the state of the measured system; and \emph{noise-noise relations}, which quantify the tradeoff between how accurately a measurement can jointly determine the values of two non-commuting observables. New measures and relations for noise and disturbance have been proposed~\cite{Ozawa2003, Busch2013}, refined~\cite{Branciard2013, Ozawa14}, and subjected to experimental tests~\cite{Erhart2012, Rozema2012, Sulyok2013, Baek2013, Kaneda2014, Ringbauer2014, Demirel2016, Ma2016, Sulyok2017}. Initially, proposed ways to quantify noise and disturbance focused on distance measures between target observables and measurements~\cite{Ozawa2003} or the associated probability distributions~\cite{Werner2004}. More recently, interest has grown in information-theoretic measures, introduced first by Buscemi \emph{et al}.~\cite{Buscemi2014}, but also in several subsequent alternative approaches~\cite{Coles2015,Baek2016,Barchielli2016,Schwonnek2016}.
A major challenge in the study of entropic measurement uncertainty relations is to determine how tight they are.  This can be difficult for even the simplest systems, as demonstrated in~\cite{Sulyok2015}, where an allegedly tight noise-disturbance relation for orthogonal qubit observables was given and tested experimentally. Subsequently, however, a counterexample was found~\cite{Branciard2016}, showing that the relation can be violated by non-projective measurements. In this article we focus on related noise-noise uncertainty relations, experimentally testing the noise-noise tradeoff for a range of (not necessarily orthogonal) Pauli observables. By implementing 4-outcome general quantum measurements we saturate tight noise-noise relations, thereby improving upon previous experiments with projective measurements~\cite{Sulyok2015}.

\section{Theoretical Framework}

To formally study measurement uncertainty relations one must define measures for two key properties of a measurement device $\mathcal{M}$ (which may in general implement an arbitrary quantum measurement with any number of outcomes): how accurately it measures a target observable $A$ (its \emph{noise}), and how much it disturbs the quantum state during measurement (the \emph{disturbance}).
Here we are interested in noise-noise uncertainly relations and therefore restrict our discussion to the former.

While several definitions of noise have previously been studied theoretically and experimentally, we utilize the information-theoretic approach of~\cite{Buscemi2014}, formulated as follows.
Let $\{\ket{a}\}_a$ be the $d$ eigenstates of the $d$-dimensional target observable $A$ and represent $\mathcal{M}$ as a positive-operator valued measure (POVM) $\mathcal{M}=\{M_m\}_m$~\cite{NielsenChuang}. 
The noise is defined in the following scenario: the eigenstates of $A$ are randomly prepared with probability $p(a)=\frac{1}{d}$ before $\mathcal{M}$ is measured, producing an outcome $m$ with probability $p(m|a)=\Tr(M_m \ket{a}\!\!\bra{a})$.
If $\mathcal{M}$ accurately measures $A$ then the value of $m$ should allow one to determine $a$; if the measurement is noisy, $m$ yields less information about $a$.
This noise is quantified in terms of the conditional Shannon entropy: 
denoting the random variables associated with $a$ and $m$ as $\mathbb{A}$ and $\mathbb{M}$, respectively, the \emph{noise} of $\mathcal{M}$ on $A$ is~\cite{Buscemi2014} 
\begin{equation}
	N(\mathcal{M},A) = H(\mathbb{A}|\mathbb{M})=-\sum_{a,m} p(a,m)\log_2 p(a|m),
	\label{eq:noiseDefn}
\end{equation}
where $p(a,m)=p(a)p(m|a)$ and $p(a|m)$ can be calculated from Bayes' theorem~\cite{footnote_AM}.

If $A$ and $B$ are two non-commuting observables, the noises $N(\mathcal{M},A)$ and $N(\mathcal{M},B)$ (defined similarly) cannot both be zero.
Subsequently, there is a tradeoff between these quantities which can be expressed by uncertainty relations, e.g.~\cite{MaassenUffink1988,Buscemi2014}
\begin{equation}\label{eq:MUreln}
	N(\mathcal{M},A)+N(\mathcal{M},B)\ge -\log_2\max_{a,b}|\!\braket{a|b}\!|^2,
\end{equation}
but such relations are often far from tight.
More comprehensively, one may look to completely characterize the set 
\begin{equation} \label{eq:RAB}
	R(A,B)=\{(N(\mathcal{M},A),N(\mathcal{M},B)): {\mathcal{M} \textrm{\,\,is a POVM}}\}
\end{equation}
of obtainable noise values.

Recently, it has been shown~\cite{Branciard2016} that for qubit measurements one has $R(A,B)={\rm{conv\,}} E(A,B)$, where $\rm{conv}$ denotes the convex hull and $E(A,B)$ is the set of obtainable entropic preparation uncertainty values for $A$ and $B$ (see~\ref{app_theory}). This relation, derived in~\cite{Abbott2016}, allows one to characterize and experimentally probe $R(A,B)$. For projective qubit measurements, it turns out that one can obtain precisely the noise values in $E(A,B)$, but (if $A$ and $B$ are such that $E(A,B)$ is non-convex) the noise values in $R(A,B)\!\setminus\! E(A,B)$ can only be obtained by non-projective measurements~\cite{Branciard2016}.

Focusing on Pauli observables $A=\vec{a}\cdot\vec{\sigma}$ and $B=\vec{b}\cdot\vec{\sigma}$ [with $\vec{a}, \vec{b}$ two unit vectors on the Bloch sphere and $\vec{\sigma}$ = $(\sigma_x,\sigma_y,\sigma_z)$], one has
\begin{equation}\label{eq:NNregionQubits}
	\hspace{-10mm} R(A,B)={\rm{conv}}\big\{(s,t) : g(s)^2 + g(t)^2 - 2|\vec{a}\cdot\vec{b}|\,g(s)\,g(t) \le 1 - (\vec{a}\cdot\vec{b})^2\big\},	
\end{equation}
where $g$ is the inverse of the binary entropy function $h(x)$ defined for $x\in [0,1]$ as
\begin{equation}
	\textstyle h(x)=-\frac{1+x}{2}\log_2\left(\frac{1+x}{2}\right)-\frac{1-x}{2}\log_2\left(\frac{1-x}{2}\right).
\end{equation}
When $|\vec{a}\cdot\vec{b}|\gtrsim0.391$, $E(A,B)$ is convex and the entire region $R(A,B)$ can be obtained by projective measurements; for $|\vec{a}\cdot\vec{b}|\lesssim 0.391$ it is non-convex~\cite{Abbott2016,Sanchez-Ruiz1998,Vicente2008} and saturating the noise-noise tradeoff requires 4-outcome POVMs~\cite{Branciard2016} (see~\ref{app_theory} for further theoretical details).

\section{Experimental Procedure}

\begin{figure*}
	\includegraphics[width=0.99\textwidth]{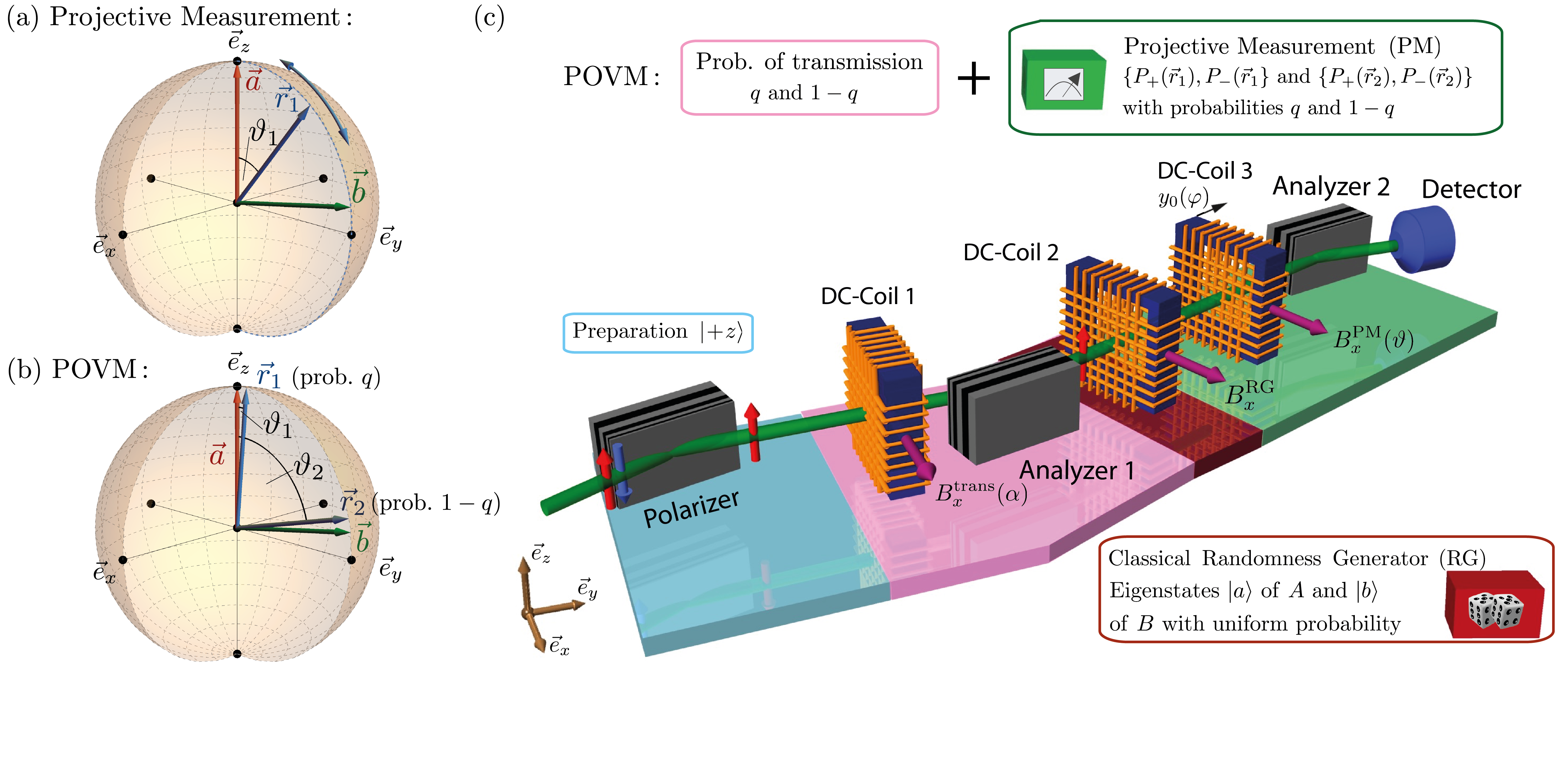}
	\caption{(a--b) \textit{Measurement strategies}. 
		The vector $\vec{r}_1(\vartheta_1,\varphi_1)$ in the Bloch sphere (a) represents a projective measurement of the observable $\vec{r}_1\cdot\vec{\sigma}$. The vectors $\vec{r}_1$ and $\vec{r}_2$ in (b) represent the projective measurements that are mixed with probabilities $q$ and $1-q$ to realize the POVM $\mathcal{M}$ of equation~\ref{eq:POVM}.
		(c) \textit{Neutron polarimeter setup of the measurement}. The red arrow indicating the state $\ket{+z}$ is rotated in DC-Coil~1 by applying a magnetic field $B_x^{\rm{trans}}(\alpha)$, before passing Analyzer~1, which projects the state onto $\ket{+z}$, with probability $q$ or $1-q$ (depending on $\alpha$). A random number generator then selects a magnetic field $B_x^{\rm{RG}}$ to apply in DC-Coil~2, which prepares one of the eigenstates $\ket{a}, \ket{b}$ of $A$ and $B$. Finally, the third magnetic field $B_x^{\rm{PM}}(\vartheta)$ in DC-Coil~3 and Analyzer~2 realize a projective measurement in the direction $\vec{r}_1$ on the neutrons passing Analyzer~1 with probability $q$, or in the direction $\vec{r}_2$ on the ensemble transmitted with probability $1-q$. The measurement direction $\vec{r}_1$ can be brought out of the $\vec{a}\vec{b}$-plane by displacing DC-Coil~3 by $y_0(\varphi)$. For further details, see~\ref{app_exp_techniques}.}
	\label{fig:SetUp}
\end{figure*}

In this work, we describe an experiment probing the noise-noise tradeoff between Pauli observables $A$ and $B$ using neutron spin qubits. Neutrons are ideal test objects for foundational experiments, since they are described by matter waves whose polarization and trajectories can be accurately manipulated and efficiently detected.

 The neutrons in our experiment are produced at the TRIGA Mark-II reactor of the Vienna University of Technology, where they are monochromatized to an average wavelength of $\lambda=2.02\,\textrm{\AA}$ and polarized by reflection on a Co-Ti supermirror. 
The particles entering the beam line are guided by a vertical magnetic field determining the quantization axis and specifying the incident spin as $\ket{+z}$, the eigenvector of $\sigma_z$.
The noise-noise tradeoff is then probed by implementing POVMs of the form
\begin{equation}
\mathcal{M} = \left\{q P_+(\vec{r}_1),q P_-(\vec{r}_1),(1-q) P_+(\vec{r}_2),(1-q)  P_-(\vec{r}_2) \right\},
\label{eq:POVM}
\end{equation} 
where $P_\pm(\vec{r}_i)=\frac{1}{2}(\id\pm\vec{r}_i\cdot\vec{\sigma})$, and the $\vec{r}_i:=\vec{r}_i(\vartheta_i,\varphi_i)$ are unit vectors on the Bloch sphere parametrized by the spherical coordinates $\vartheta_i,\varphi_i$.   
For $q=1$, equation~\ref{eq:POVM} reduces to a single projective spin measurement along the direction $\vec r_1$, see figure~\ref{fig:SetUp}(a), while for a value of $q$ between $0$ and $1$ it corresponds to a mixture of projective measurements along the directions $\vec{r}_1$ and $\vec{r}_2$ with probabilities $q$ and $1-q$, see figure~\ref{fig:SetUp}(b).

To perform the required measurements, an experimental setup [see figure~\ref{fig:SetUp}(c)] similar to that in Ref.~\cite{Sulyok2013} is employed. The individual elements of the POVM are successively measured allowing the statistics for the whole POVM to be reconstructed. To this end, the initial spins are first rotated by DC-Coil~1 before being transmitted through Analyzer~1 with probabilities $q,1-q$ depending on the incident angle of spin states. After Analyzer~1, one of the observables' eigenstates $\ket{a}$ or $\ket{b}$ (with eigenvalues $a,b=\pm 1$) is generated uniformly at random by inducing an appropriately chosen rotation at DC-Coil~2. DC-Coil~3 is set so that the incoming neutrons pass Analyzer~2 with probabilities $q\Tr[P_\pm(\vec{r}_1)\ket{a}\!\!\bra{a}]$ or $(1-q)\Tr[P_\pm(\vec{r}_2)\ket{a}\!\!\bra{a}]$, and  likewise for the $\ket{b}$ eigenstates. At the end of the beam line a boron trifluoride detector registers all incoming neutrons so that, given these settings, one of $q P_\pm(\vec{r}_1)$ or $(1-q)P_\pm(\vec{r}_2)$ is measured; each detection for one of these 4 measurement operators corresponds to a different outcome $m$. We thereby record the counts $I_{a,m}$ of measuring outcome $m$ when $\ket{a}$ is prepared, and similarly for $I_{b,m}$, and estimate the probabilities $p(a,m)=I_{a,m}/\sum_{a,m}I_{a,m} $ and $p(b,m)=I_{b,m}/\sum_{b,m}I_{b,m}$ (under a standard fair-sampling assumption), permitting the noises $N(\mathcal{M},A)$ and $N(\mathcal{M},B)$ to be calculated from equation~\ref{eq:noiseDefn}.

\subsection*{Results}

\begin{figure*}
	\includegraphics[width=1.0\textwidth]{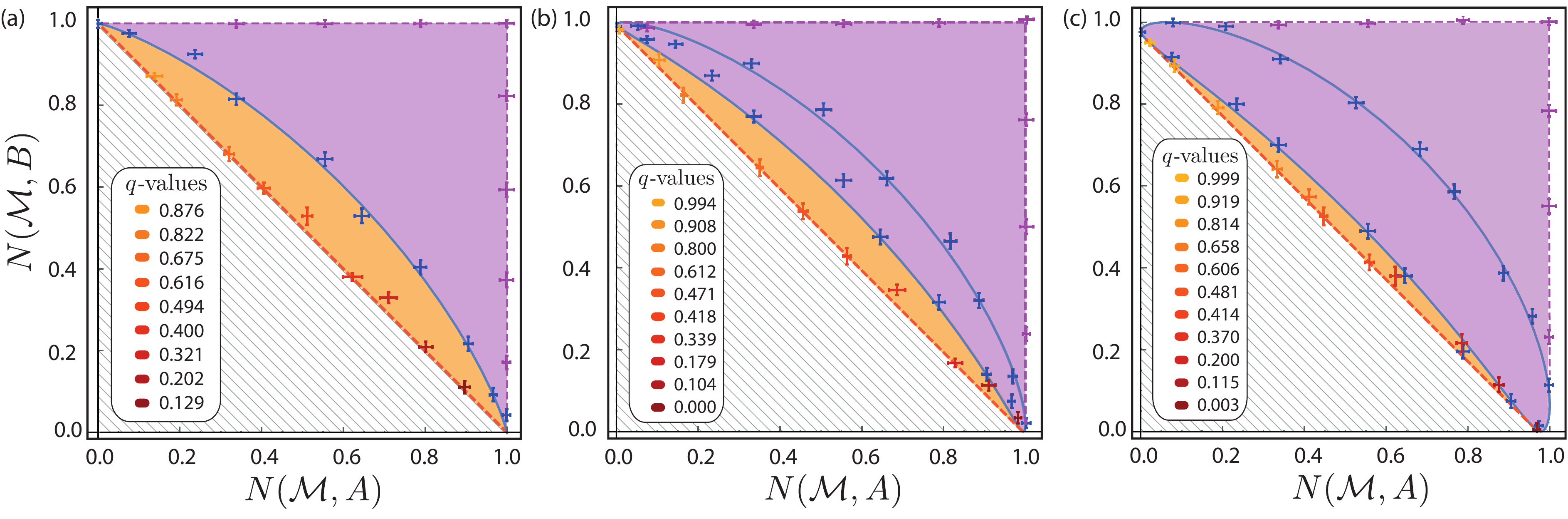}
	\caption{Plots of the noise-noise regions $R(A,B)$ for $A=\vec{a}\cdot\vec{\sigma},$ $B= \vec{b}\cdot\vec{\sigma}$ with (a) $\vec{a}\cdot\vec{b}\simeq0$, (b) $\vec{a}\cdot\vec{b} \simeq 0.07$ and (c) $\vec{a}\cdot\vec{b}\simeq0.19$. The shaded noise-noise regions are separated here into two areas. The purple area shows the region $E(A,B)$ reachable by projective measurements; the blue data points are measured in the $\vec{a}\vec{b}$-plane ($\varphi_1 = \frac{\pi}{2}$) starting from $\vartheta_1=0$ ($\vec{r}_1=\vec{a}$) and increasing in steps of $\Delta\vartheta_1=10^\circ$. When $\vec{a}\cdot\vec{b}=0$ the noise is symmetric around $\vartheta_1=\pi/2$; otherwise, the closed blue curve is obtained. The purple points are obtained by taking $q=1$ and rotating $\vec{r}_1=\vec{a}$ (top boundary) and $\vec{r}_1=\vec{b}$ (right boundary) out of the $\vec{a}\vec{b}$-plane by increasing the azimuthal angle $\varphi_1$. The orange area corresponds to $R(A,B)\setminus E(A,B)$, and the noise-noise values inside it can only be reached by POVMs; the points on its lower-left linear boundary can be obtained by a 4-outcome POVM realized as a mixture of two projective measurements along some fixed directions $\vec{r}_1$ and $\vec{r}_2$, with varying values of $q$, see equation~\ref{eq:POVM} and figure~\ref{fig:SetUp}(b). Outside of these regions, the values of $(N(\mathcal{M},A),N(\mathcal{M},B))$ in the hatched areas are forbidden and cannot be reached by any quantum measurement. Error bars correspond to one standard deviation arising from the Poissonian statistics of the neutron count rate. 
	}
	\label{fig:PlotResults}
\end{figure*}

\begin{figure*}[!ht]
\begin{center}
	\includegraphics[width=.7\textwidth]{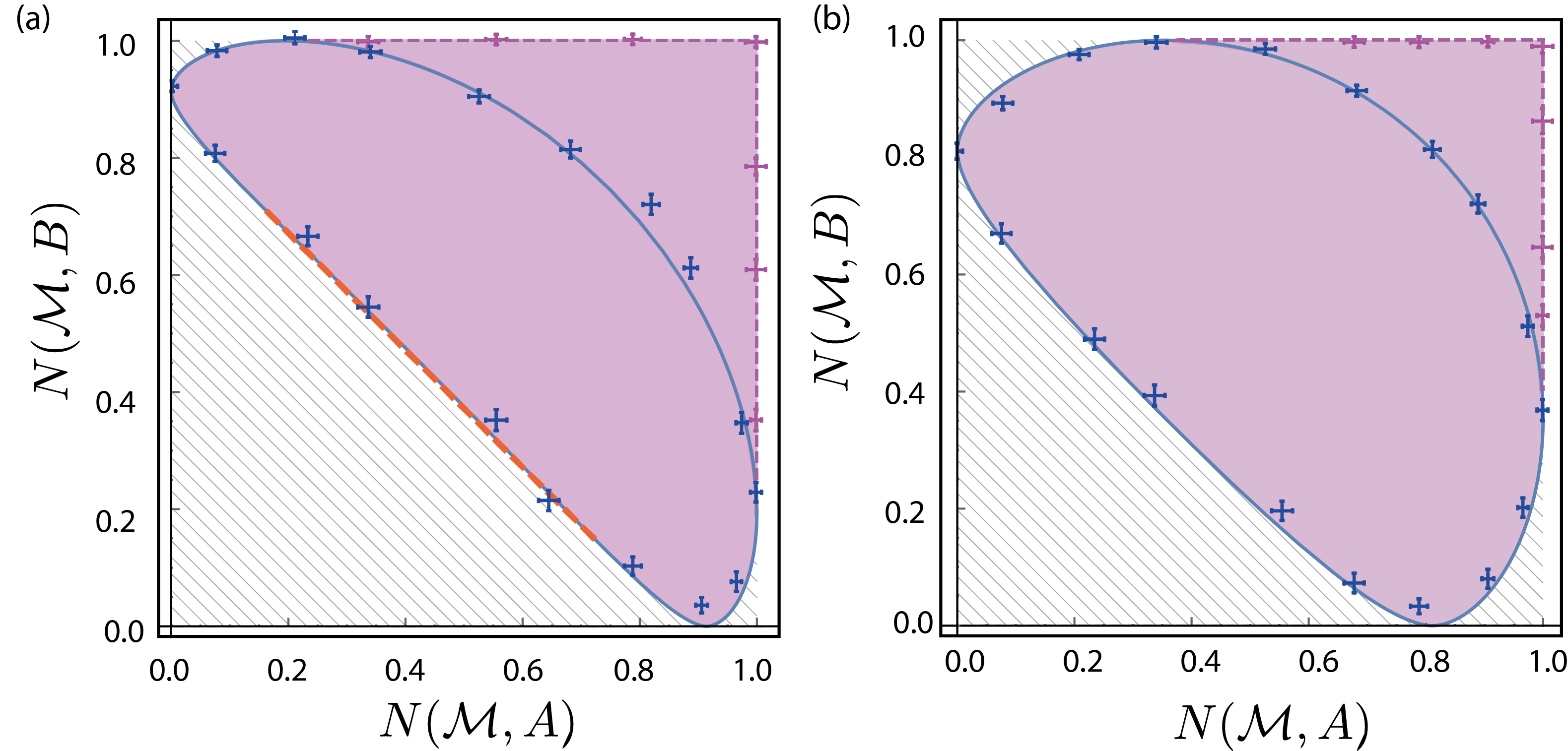}
	\caption{Plots of the noise-noise regions $R(A,B)$ with (a) $\vec{a}\cdot\vec{b}\simeq 0.35$, and (b) $\vec{a}\cdot\vec{b}\simeq 0.5$. These cases are on either side of the value $\vec{a}\cdot\vec{b}\simeq 0.391$ at which $E(A,B)$ becomes convex. In (a), due to the size of the error bars (as specified in figure~\ref{fig:PlotResults}), improvements beyond projective measurements to obtain noise values on the orange dashed line are experimentally no longer possible, while in (b) the optimal theoretical tradeoff is  already attained with projective measurements.}
	\label{fig:PlotResults2}
\end{center}
\end{figure*}

To obtain all the relevant results the following procedure is applied. We take $\vec{a}=\vec{e}_z$ (the unit vector in the $z$ direction) and choose $\vec{b}$ in the $yz$-plane, thus determining the value $|\vec{a}\cdot\vec{b}|$ characterizing the noise-noise tradeoff. We initially take $q=1$ so that the projectors $P_{\pm}(\vec{r}_1)$ are measured on the entire neutron ensemble. The vector $\vec{r}_1=\cos(\vartheta_1)\vec{e}_z+\sin(\vartheta_1)\vec{e}_y$ is rotated in the interval $\vartheta_1 \in[0^\circ,180^\circ]$ with increments of $\Delta \vartheta_1\simeq10^\circ$ [see figure~\ref{fig:SetUp}(a)].  The variation of the polar angle changes the probabilities of passing Analyzers 1 and 2 and reaching the detector, and thus of $p(a|m)$ and $p(b|m)$. When $\vartheta_1=0$ the projectors are $P_\pm(\vec{r}_1)=P_\pm(\vec{a})$ and the probability $p(a|m)$ is maximally peaked, while $p(b|m)$ is evenly distributed. For $\vec{r}_1=\vec{b}$ the situation is exactly reversed. When $\vec r_1$ is in between $\vec a$ and $\vec b$, these measurements attain the lower-left boundary of $E(A,B)$ and are optimal amongst projective measurements. The upper-right boundaries of $E(A,B)$ can, for completeness, be obtained by rotating $\vec{r}_1$ out of the plane spanned by $\vec{a}$ and $\vec{b}$ by an azimuthal angle $\varphi_1$ (varied experimentally by displacing DC-Coil~3, see figure~\ref{fig:SetUp}(c)), increasing the noise with respect to both $A$ and $B$ (Figs.~\ref{fig:PlotResults} and~\ref{fig:PlotResults2}). 

For $|\vec{a}\cdot\vec{b}|\gtrsim 0.391$ projective measurements are optimal and this approach saturates the noise-noise tradeoff. This is no longer true for $|\vec{a}\cdot\vec{b}|\lesssim 0.391$ and the noise may be decreased further by non-projective measurements. To saturate the tradeoff and attain the lower-left boundary of $R(A,B)$ the mixing parameter $q$ is varied to implement the full 4-outcome POVM $\mathcal{M}$.
This is done by mixing the statistics obtained by the projectors $P_\pm(\vec{r}_1)$ and $P_\pm(\vec{r}_2)$, where the polar angles $\vartheta_1$ and $\vartheta_2$, associated with $\vec{r}_1, \vec{r}_2$, are determined by the \emph{projective} measurements $\mathcal{M}_i$ minimizing $N(\mathcal{M}_i,A)+N(\mathcal{M}_i,B)$. With these angles fixed (giving vectors $\vec r_1$ and $\vec r_2$ between $\vec a$ and $\vec b$, symmetric about their angle bisector; see figure~\ref{fig:SetUp}(b)), a range of POVMs are implemented by varying $q$, i.e., changing the ratio of transmitted neutrons in Analyzer~1, by $\Delta q\simeq 0.1$.
These measurements attain the boundary of the orange noise-noise region $R(A,B)$ in figure~\ref{fig:PlotResults}. 

The measurement results for three different vectors $\vec{b}$ (for which $E(A,B)$ is non-convex) are given in figure~\ref{fig:PlotResults}, with (a) $\vec{a}\cdot\vec{b}\simeq 0$, (b) $\vec{a}\cdot\vec{b} \simeq 0.07$ and (c) $\vec{a}\cdot \vec{b}\simeq 0.19$. The noise-noise region $R(A,B)$ is broken into two subregions: the purple region $E(A,B)$ of values attainable with projective measurements, and the orange region $R(A,B)\!\setminus\!E(A,B)$. The closed blue curve shows the values attainable by projective measurements in the $\vec{a}\vec{b}$-plane, while the dashed orange line shows the optimal values attainable with POVMs. In figure~\ref{fig:PlotResults}(a), $\vec{a}$ is perpendicular to $\vec{b}$ and projective measurements $\mathcal{M}$ in the $\vec{a}\vec{b}$-plane give noise values that lie on the lower-left boundary (blue curve) of $E(A,B)$. To saturate the noise-noise tradeoff, we mix projective measurements in directions $\vec{r}_1=\vec{a}$ and $\vec{r}_2=\vec{b}$. The resulting points are color coded from red ($q=0$) to orange ($q=1$). We see that POVMs give a considerable improvement on the uncertainty relation over projective measurements, which previous experiments had been restricted to~\cite{Sulyok2015}. For instance, for the data point corresponding to $q \simeq 0.494$ on figure~\ref{fig:PlotResults}(a), we obtain noise values of $(N(\mathcal{M},A),N(\mathcal{M},B)) = (0.511\pm 0.012,0.529 \pm 0.021)$. This violates the relation $g(N(\mathcal{M},A))^2 +  g(N(\mathcal{M},B))^2 \le 1$ satisfied by all projective measurements (corresponding to the lower boundary of the purple region $E(A,B)$, see~\ref{app_theory}), by more than 6 standard deviations. When the eigenstates of $B$ approach those of $A$, as is the case in figure~\ref{fig:PlotResults}(b), the lower boundary of $R(A,B)$ (orange) and, more noticeably the purple region $E(A,B)$ start shifting downwards. This becomes more apparent in figure~\ref{fig:PlotResults}(c), where the optimal choice of $\vec{r}_1$ and $\vec{r}_2$ to mix is obtained for the projective measurements giving $(N(\mathcal{M},A),N(\mathcal{M},B))\simeq(0.02,0.95)$ and $(N(\mathcal{M},A),N(\mathcal{M},B))\simeq (0.95,0.02)$, respectively, corresponding to $\vartheta_1\simeq 5^\circ$ and $\vartheta_2 \simeq 74^\circ$ ($\arccos (\vec{a}\cdot\vec{b})\simeq 79^\circ$). By realizing the POVM accordingly for a range of $q$ values we again succeed in saturating the noise-noise tradeoff. (See~\ref{app_data} for further quantitative details of the measurements.)

In figure~\ref{fig:PlotResults2} we present two cases with inner products (a) $\vec{a}\cdot\vec{b}\simeq 0.35$ and (b) $\vec{a}\cdot\vec{b}\simeq 0.5$, on either side of the critical value of $|\vec{a}\cdot\vec{b}|\simeq 0.391$ at which the region $E(A,B)$ becomes convex. In figure~\ref{fig:PlotResults2}(a) the dashed orange line from $(N(\mathcal{M},A),N(\mathcal{M},B))\simeq (0.17,0.70)$ to $(0.70,0.17)$ implies that projective measurements are theoretically incapable of saturating the noise-noise tradeoff, but improvements by POVMs are no longer resolvable in our experiment. In figure~\ref{fig:PlotResults2}(b) the region $R(A,B)$ is already convex and can be fully attained with projective measurements, hence improvements by general POVMs are no longer possible.

\section{Discussion}
A measurement device cannot jointly measure two non-commuting observables with arbitrary precision, and thus there is a tradeoff between the accuracy with which they can both be measured, captured by noise-noise uncertainty relations. Using a definition of noise that quantifies how well a measurement device can distinguish eigenstates of non-commuting observables~\cite{Buscemi2014}, we experimentally tested tight entropic noise-noise uncertainty relations for qubits~\cite{Branciard2016} for various pairs of Pauli spin observables. For closely aligned observables, we saw that the uncertainty relation could be saturated with simple projective measurements.
However, we verified experimentally that this is not generally the case and that four-outcome POVMs yield better measurement results that saturate the uncertainty relation when projective measurements cannot. It is interesting to note that advantages accorded by POVMs over projective measurements have also been reported for other features of measurements~\cite{Holevo1973}, and it would be interesting to clarify this connection further in the future.
Our study, which focused on noise-noise relations, paves the way for further experiments testing entropic noise-\emph{disturbance} relations~\cite{Buscemi2014}. For such relations on qubit systems general quantum measurements again offer advantages over projective measurements, but their experimental realization necessitates the implementation of nontrivial post-measurement transformations on the measured states~\cite{Branciard2016}. 

\ack
B.D., S.S.\ and Y.H.\ acknowledge support by the Austrian science fund (FWF) Projects No.\ P30677-N20 and No.\ P27666-N20. A.A.\ and C.B.\ acknowledge financial support from the “Retour Post-Doctorants” program (ANR-13-PDOC-0026) of the French National Research Agency.

\appendix

\section{Theoretical framework}
\label{app_theory}

Let us first give a more detailed presentation of the framework in which the information-theoretical noise is defined, as well as on the form of the regions $R(A,B)$ and $E(A,B)$. Further details can be found in~\cite{Buscemi2014,Branciard2016}.

\subsection{Operational scenario defining information-theoretic noise}

The most general model of a quantum measurement is that of a \emph{quantum instrument}, which completely describes both the statistics of a measurement and the transformation it induces on the measured system.
To define the noise, however, only the statistics (and not the transformation) are of interest to us, and these can be described by a \emph{positive-operator valued measure (POVM)}.
Recall that a POVM $\mathcal{M}$ is a collection $\{M_m\}_m$ of Hermitian positive semidefinite operators $M_m$ satisfying $\sum_m M_m=\id$, where $\id$ is the identity operator.
For a given quantum state $\rho$, the probability of obtaining outcome $m$ is $\Tr[M_m \rho]$.\\

\begin{figure}
\begin{center}
	\includegraphics[width=0.65\textwidth]{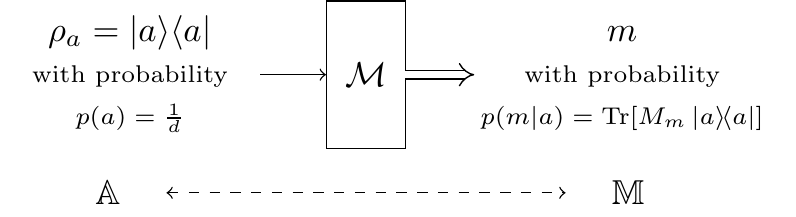}
	\caption{A schematic of the operational scenario defining the noise $N(\mathcal{M},A)$ of a measurement $\mathcal{M}$ with respect to the target observable $A$~\cite{Branciard2016}. The eigenstates $\ket{a}$ of $A$ are prepared uniformly at random before being measured by $\mathcal{M}$, which produces an outcome $m$.}
	\label{fig:framework}
\end{center}
\end{figure}

The information-theoretic definition of noise $N(\mathcal{M},A)$ is best understood in the operational framework described in the main text, and illustrated in figure~\ref{fig:framework} below.
For simplicity, let $A$ be a $d$-dimensional non-degenerate observable with eigenstates $\{\ket{a}\}_a$.
The operational scenario can be seen as an experiment in which the eigenstates $\ket{a}$ are prepared uniformly at random, i.e., with probability $p(a)=\frac{1}{d}$, before being measured by $\mathcal{M}$.
The result of the measurement is the outcome $m$ with probability $p(m|a)=\Tr[M_m \ket{a}\!\!\bra{a}]$; note that a non-projective measurement may have more than $d$ outcomes.
One thus has the joint distribution $p(a,m)=p(a)p(m|a)=\frac{1}{d}\Tr[M_m \ket{a}\!\!\bra{a}]$ specifying the probability of preparing $\ket{a}$ and obtaining outcome $m$.
It will be convenient to denote the random variables corresponding to $a$ and $m$ by $\mathbb{A}$ and $\mathbb{M}$, respectively, where we use the double-struck letters to differentiate the classical random variable $\mathbb{A}$ from the quantum observable $A$.

Given a particular measurement outcome $m$ one may ask what state $\ket{a}$ was prepared. 
If the measurement is noiseless, one should be able to determine this with certainty; conversely, the uncertainty in which eigenstate of $A$ was prepared, given $m$, is used to quantify the noise of $\mathcal{M}$ with respect to $A$.
More precisely, this is quantified via the conditional Shannon entropy as~\cite{Branciard2016}
\begin{equation}\label{eq:noiseDef}
	N(\mathcal{M},A)=H(\mathbb{A}|\mathbb{M})=-\sum_{a,m}p(a,m)\log_2 p(a|m),
\end{equation}
where
\begin{equation}\label{eq:condProb}
	p(a|m)=\frac{p(a,m)}{\sum_a p(a,m)}=\Tr\left[\ket{a}\!\!\bra{a}\frac{M_m }{\Tr[M_m]}\right].
\end{equation}
A large conditional Shannon entropy $H(\mathbb{A}|\mathbb{M})$ means there is a lot of uncertainty in the value of $a$ (i.e., the eigenstate prepared) given an observation $m$, so this definition indeed quantifies the intuitive notion of noise discussed above.\\

The noise $N(\mathcal{M},A)$ can thus be easily measured by preparing randomly the eigenstates $\ket{a}$ of $A$ before measuring them and estimating the joint distribution $p(a,m)$ from the observed incident counts.
To probe the noise-noise tradeoff, $N(\mathcal{M},A)$ and $N(\mathcal{M},B)$ must both be calculated, which requires performing two such experiments (preparing randomly the states $\ket{b}$ in the second).
In practice, both experiments can be performed simultaneously by preparing the eigenstates $\{\ket{a},\ket{b}\}_{a,b}$ at random with probability $\frac{1}{2d}$ and separating out the statistics $p(a,m)$ and $p(b,m)$; this is precisely what we do in the experiment described in the main text.

\subsection{$R(A,B)$ and its connection with entropic preparation uncertainty}

The set $R(A,B)=\big\{\big(N(\mathcal{M},A),N(\mathcal{M},B)\big):\mathcal{M} \textrm{ is a valid POVM}\big\}$ of obtainable noise values completely characterizes the noise-noise tradeoff relation.
Characterizing $R(A,B)$ is, in general, difficult due to the nonlinearity of~\ref{eq:noiseDef} and the need to consider the noise obtainable by arbitrary POVMs (which themselves are not easily characterized beyond the simplest systems)~\cite{Buscemi2014,Branciard2016}.
To do so for qubit measurements, we exploit a relation to entropic \emph{preparation} uncertainty relations.
Let $H(A|\rho)$ be the measurement entropy of $A$ for a state $\rho$, defined as
\begin{equation}
	H(A|\rho)=-\sum_a \Tr\big[\ket{a}\!\!\bra{a}\rho\big]\log_2 \Tr\big[\ket{a}\!\!\bra{a}\rho\big].
\end{equation}
The entropic preparation region
\begin{equation}
	E(A,B)=\left\{\big(H(A|\rho),H(B|\rho)\big) : \rho \textrm{ is a density matrix}\right\}
\end{equation}
characterizes how well-defined the values of $A$ and $B$ can be for any quantum state $\rho$, and is a key object in the study of entropic preparation uncertainty relations~\cite{Abbott2016}.\\

In \cite{Branciard2016} it was shown that $R(A,B)\subseteq {\rm{conv\,}} E(A,B)$, where ${\rm{conv}} $ denotes the convex hull.
Moreover, the authors showed that one has \emph{equality} for qubit systems, and for such systems $E(A,B)$ is well-understood.
Indeed, it has been shown that for Pauli observables $A=\vec{a}\cdot\vec{\sigma}$ and $B=\vec{b}\cdot\vec{\sigma}$~\cite{Abbott2016}
\begin{equation}
	E(A,B)=\left\{(s,t) : g(s)^2 + g(t)^2 - 2|\vec{a}\cdot\vec{b}|\,g(s)\,g(t) \le 1 - (\vec{a}\cdot\vec{b})^2 \right\}
\end{equation}
from which one obtains equation~\ref{eq:NNregionQubits} of the main text (after which $g$ is defined).
By exploiting the fact that $E(A,B)$ is convex when $|\vec{a}\cdot\vec{b}|\gtrsim0.391$, for which one thus has $R(A,B)=E(A,B)$, one can, for such observables, write the explicit tight noise-noise uncertainty relation
\begin{equation} \label{eq:ineqExplicit}
	\hspace{-20mm} g\big(N(\mathcal{M},A)\big)^2 + g\big(N(\mathcal{M},B)\big)^2 - 2|\vec{a}\cdot\vec{b}|\,g\big(N(\mathcal{M},A)\big)\,g\big(N(\mathcal{M},B)\big) \le 1 - (\vec{a}\cdot\vec{b})^2. \quad
\end{equation}
When $|\vec{a}\cdot\vec{b}|\lesssim0.391$ it is not possible to have an explicit \emph{inequality} in this way. 
Nonetheless, the boundary $E(A,B)$ can be found and expressed in a piecewise form.
Indeed, note  that only the ``lower boundary'' of $E(A,B)$ (i.e., the points $(s,t)$ on the boundary of $E(A,B)$ for which there are no points $(u,v)$ in $E(A,B)$ with $u<s$ or $v<t$) is non-convex for $|\vec{a}\cdot\vec{b}|\lesssim0.391$ (see figure~\ref{fig:PlotResults} of the main text).
The convex hull of $E(A,B)$ can be readily computed numerically and one thus obtains a linear boundary between two points $\big(N^*_1(\mathcal{M},A),N^*_1(\mathcal{M},B)\big)$ and $\big(N^*_2(\mathcal{M},A),N^*_2(\mathcal{M},B)\big)$ (corresponding to the two points obtained by the projective measurements that must be mixed to saturate the lower boundary of $R(A,B)$) and the curve given by the points saturating equation~\ref{eq:ineqExplicit} elsewhere.

For orthogonal Pauli measurements $\vec{a}\cdot\vec{b}=0$, it is worth noting that one has simply the tight inequality
\begin{equation}
	N(\mathcal{M},A)+N(\mathcal{M},B) \ge 1,
\end{equation}
which takes precisely the same form as the Maassen and Uffink inequality (see equation~\ref{eq:MUreln} in the main text).
In contrast, \emph{projective} measurements (which can only give points in $E(A,B)$) satisfy the relation 
\begin{equation}
	g\big(N(\mathcal{M},A)\big)^2+g\big(N(\mathcal{M},B)\big)^2 \le 1,
\end{equation}
which can therefore be violated by POVMs.\\

Experimentally, we are interested in saturating the noise-noise tradeoff relation, achievable by performing measurements for which $\big(N(\mathcal{M},A),N(\mathcal{M},B)\big)$ is on the boundary of $R(A,B)$.
Of particular interest is the ``lower boundary'' of $R(A,B)$; measurements obtaining points on this are optimal with respect to the noise-noise tradeoff.
In order to obtain such points when $E(A,B)$ is non-convex, one must consider non-projective measurements.
In particular, to this end we implement 4-outcome POVMs which correspond to probabilistic mixtures of projective measurements (and recording which projective measurement was performed).
Projective measurements suffice to obtain all points in $E(A,B)$ (and thus in $R(A,B)$ with $E(A,B)$ is already convex).

\section{Experimental techniques}
\label{app_exp_techniques}

The experimental approach we use to probe uncertainty relations with neutron spins is similar to that used in Ref.~\cite{Sulyok2013}, and further details on the general approach can be found therein. The primary additional challenge in the present experiment is to implement the 4-outcome POVMs needed to probe the tight entropic uncertainty relation.

The initially polarized neutrons encounter two supermirror analyzers as shown in figure~\ref{fig:SetUp}(c) of the main text, which separate the neutrons stochastically according to their up and down spins by reflection on a magnetic multilayer structure, with the transmitted beams continuing to the next stage of the experiment. The polarizer functions as a Stern-Gerlach magnet with a similar working principle. Each supermirror has a probability of transmitting the neutrons depending on the relative angle between the incident spin and the analyzer orientation. The orientations of the analyzers are kept static along the positive $z$-direction (aligned with $\ket{+z}$) while the incoming neutron spin state is changed dynamically by the first and third DC-coils. 

The three DC-coils in the setup all work identically. Static magnetic fields are generated by direct currents fed into wires arranged as solenoids. One wire spirals helically along the vertical axis and another wire winds around the $x$-axis perpendicular to the neutron's $y$-direction of propagation. The large dimensions of the coils guarantee that the magnetic field inside the solenoids can be regarded as homogeneous for the neutrons. The purpose of the vertical magnetic field is to compensate for the exterior guide field and Earth's magnetic field, which are effectively nullified in the solenoids.

The purpose of the lateral coil is to induce a unitary Larmor precession of the initial spin. Classically, the field in the solenoid exerts a torque on the polarization vector, which quantum-mechanically is described by the unitary operator
\begin{equation}
U(\alpha)=\exp\left(i \frac{\alpha}{2} \sigma_x\right) = \exp\left(i \frac{\gamma B_x t}{2} \sigma_x\right),
\end{equation}
where the rotation angle $\alpha = \gamma B_x t$ is defined by the magnitude of the magnetic field $B_x$ in \textit{x}-direction, $\gamma$ being the neutron gyromagnetic factor $\gamma$~=~$1.833\times10^8~\textrm{rad/(s T)}$, and the time of flight through the solenoid $t =\frac{l}{v_n} \cong \frac{0.019~\textrm{m}}{1958~\textrm{m}/\textrm{s}}~\approx 10~\mu\textrm{s}$. The rotation angle $\alpha$ is controlled by the current that generates the magnetic field.\\

Let $\ket{\psi} \equiv U_{1}(\alpha)\ket{+z} = \exp(i \frac{\alpha}{2} \sigma_x)\ket{+z}$ be the state prior to Analyzer~1 and $\ket{-\psi} \equiv U_{1}(\alpha+\pi)\ket{+z} = \exp(i \frac{\alpha+\pi}{2} \sigma_x)\ket{+z}$ be the orthogonal spin state.  
The ideal approach to implementing the 4-outcome POVM in equation~\ref{eq:POVM} would be to perform the projective measurement $\{P_\pm(\vec{r}_1)\}_\pm$ on the sub-ensemble of neutrons transmitted (with probability $q$\,=\,$|\braket{+z|\psi}|^2 = \cos^2\left(\frac{\alpha}{2}\right)$) by Analyzer~1 and the projective measurement $\{P_\pm(\vec{r}_2)\}_\pm$ on the reflected  sub-ensemble (i.e., with probability $1-q$\,=\,$|\braket{+z|-\psi}|^2 = \sin^2\left(\frac{\alpha}{2}\right)$). 
In practice, instead of implementing four separate beams and detectors, at each analyzer the reflected parts are discarded and the four operators are measured sequentially by applying the appropriate rotations at DC-Coils 1 and 3. 
These four configurations, corresponding to the four POVM elements, along with the randomly chosen state to prepare ($\ket{\pm a}$ or $\ket{\pm b}$) are thus cycled in 60 second slots while keeping the total beam intensity \emph{de facto} constant.
The counts $I_{a,m}$ and $I_{b,m}$ for each combination of state preparation ($\pm a$ or $\pm b$) and outcome $m$ are thereby obtained and recorded.

Experimentally, this means that in order to measure the POVM elements $q P_\pm(\vec{r}_1)$, $\alpha$ is chosen so that $|\braket{+z|\psi}|^2=q$ and DC-Coil~3 plus Analyzer~2 are conditioned to measure $P_\pm(\vec{r}_1)=\ket{\pm\vec{r}_1}\!\!\bra{\pm\vec{r}_1}$, where $\ket{+\vec{r}_1} =  U_{3}(\vartheta_1,\varphi_1)\ket{+z}$ and $\ket{-\vec{r}_1} =  U_{3}(\vartheta_1+\pi,-\varphi_1)\ket{+z}$ (since DC-Coil~3 controls the angle of the neutrons before Analyzer~2, and thus both the direction of the projection and which outcome $\pm$ is measured).
To measure $(1-q)P_\pm(\vec{r}_2)$, $\alpha$ is changed to $\alpha+\pi$ so that $|\braket{+z|\psi}|^2=1-q$ and DC-Coil~3 plus Analyzer~2 measure instead $P_\pm(\vec{r}_2)=\ket{\pm\vec{r}_2}\!\!\bra{\pm\vec{r}_2}$ (with $\ket{+\vec{r}_2}=U_{3}(\vartheta_2,\varphi_2)\ket{+z}$ and $\ket{+\vec{r}_2}=U_{3}(\vartheta_2+\pi,-\varphi_2)\ket{+z}$).
The preparation of the desired state is effectuated by a rotation induced by DC-Coil~2, which is randomly chosen based on the signal from a uniform random number generator.
As described in the main text, $\ket{\pm a}$ correspond to directions $\pm \vec{e}_z$, while $\ket{\pm b}$ are chosen in the $yz$-plane.

After monochromatization and polarization the neutron count rate at the tangential beam port is roughly $2000\, \textrm{cm}^{-2}\, \textrm{s}^{-1}$.
The count rate at the last detector, which is $\approx$~3\,m from the polarizer, is approximately 40 neutrons per second at maximum, which is affected by the beam divergence of approximately $1^{\circ}$, the transmission efficiency of the supermirrors ($40\,\%$) and the scattering and absorption of neutrons in the copper wires of the coils. 
The detection efficiency for thermal neutrons is almost 1, owing to the high absorption cross section of ~$\!^{10}$B enriched BF$_3$ gas detector (cylindrical counter tube with 6\,cm opening diameter and 40\,cm length). The discrete counts at fixed rates in time are described by a Poisson distribution, which implies that one standard deviation of statistical error is given by the square root of the mean value.
Depolarization through ambient magnetic fields is suppressed by a 13 Gauss magnetic guide field. 
A small imperfect spin separation in the supermirror leads to a slight mixture of spin states and therefore to a loss of contrast from 100\,\% to roughly 98\,\%. To cope with this systematic imperfection, the intensity modulation of the polarization is fitted with $\frac{1}{2}(c+d\cos(x))$ which, in the ideal case would simply be $\frac{1}{2}(1+\cos(x))$. In order to take the efficiency of the detector into account, not the absolute, but the relative values of the fit parameters $c,d$ are used. 

\section{Additional details on the data evaluation}
\label{app_data}

For each pair of measurements $A$ and $B$ for which the uncertainty relation was to be probed, a range of different measurements $\mathcal{M}$ were implemented using the experimental methods described in the previous section (each giving one point on figure~\ref{fig:PlotResults} or figure~\ref{fig:PlotResults2} of the main text).
For each such measurement (and choice of observables) the counts $I_{a,m}$ and $I_{b,m}$ are obtained, where $m=1,2,3,4$ (corresponding to the outcomes of $\mathcal{M} = \left\{q P_+(\vec{r}_1),q P_-(\vec{r}_1),(1-q) P_+(\vec{r}_2),(1-q)  P_-(\vec{r}_2) \right\}$), giving a total of $16$ counts.

In order to calculate the noises $N(\mathcal{M},A)$ and $N(\mathcal{M},B)$ from these counts, the joint probability distributions $p(a,m)$ and $p(b,m)$ are first estimated as
\begin{equation}\label{eq:jointDistFromCnts}
	p(a,m) = \frac{I_{a,m}}{\sum_{a,m}I_{a,m}}~, \qquad p(b,m) = \frac{I_{b,m}}{\sum_{b,m}I_{b,m}}\,.
\end{equation}
From equation~\ref{eq:condProb} one can calculate that the conditional probabilities $p(a|m)$ and $p(b|m)$ are thus given by
\begin{equation}
	p(a|m) = \frac{I_{a,m}}{\sum_a I_{a,m}}~, \qquad p(b|m) = \frac{I_{b,m}}{\sum_b I_{b,m}}
\end{equation}
from which the noise can be be calculated directly from equation~\ref{eq:noiseDef}.\\

In order to saturate the noise-noise uncertainty relation with POVMs, we are particularly interested in families of POVMs 
\begin{equation}
\mathcal{M}=\left\{q P_+(\vec{r}_1),q P_-(\vec{r}_1),(1-q) P_+(\vec{r}_2),(1-q)  P_-(\vec{r}_2) \right\}
\end{equation}
 with different values of $q$ but $\vec{r}_1$ and $\vec{r}_2$ fixed.
While the target value of $q$ is chosen, as described earlier, by controlling the current in DC-Coil~1, in practice the effective value of $q$ might vary slightly from the desired one.
This is a consequence of the application of high currents in the wires which cause slight variations of resistance, leading to fluctuations of the magnetic field, over time.
More precise estimates of the effective values of $q$ implemented can be calculated from the counts $I_{a,m}$ and $I_{b,m}$.
To this end, note that $p(m=1)+p(m=2)=q$; in practice, one may arrive at slightly different values by calculating this from $p(a,m)$ or $p(b,m)$ due to statistical fluctuations, so we estimate $q$ as the average obtained from both these distributions.
We thus have 
\begin{eqnarray}\label{eq:qCalc}
	q &= \frac{1}{2}\left( \sum_a \big[p(a,m=1)+p(a,m=2)\big] + \sum_b \big[p(b,m=1)+p(b,m=2)\big] \right)\nonumber\\
	 &= \frac{1}{2}\left( \frac{\sum_a(I_{a,1} + I_{a,2})}{\sum_{a,m}I_{a,m}} + \frac{\sum_b(I_{b,1} + I_{b,2})}{\sum_{b,m}I_{b,m}} \right).
\end{eqnarray}
Note that this value of $q$ is only used to help present and understand our results for families of measurements with varying values of $q$, but is not needed to calculate the noises which are computed directly from the counts $I_{a,m}$ and $I_{b,m}$.

\begin{figure}[!ht]
	\begin{subfigure}[c]{1.0\textwidth}		
		\includegraphics[width=0.93\textwidth]{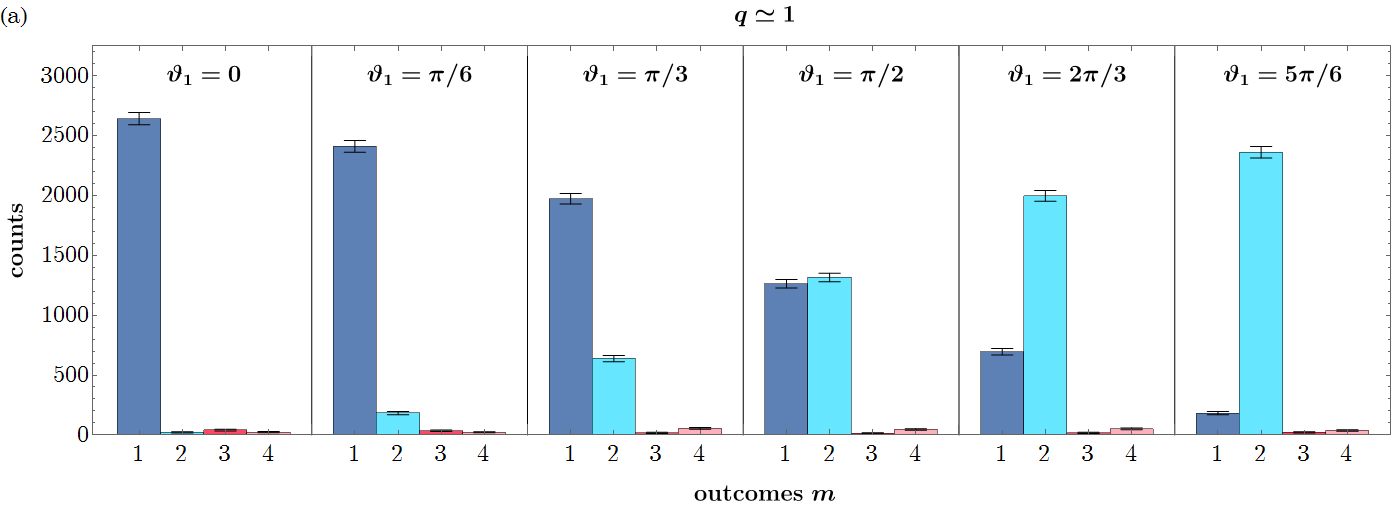}\phantomsubcaption\label{fig:AnoiseProj}\newline
	\end{subfigure}
	\begin{subfigure}[c]{1.0\textwidth}
		\includegraphics[width=0.93\textwidth]{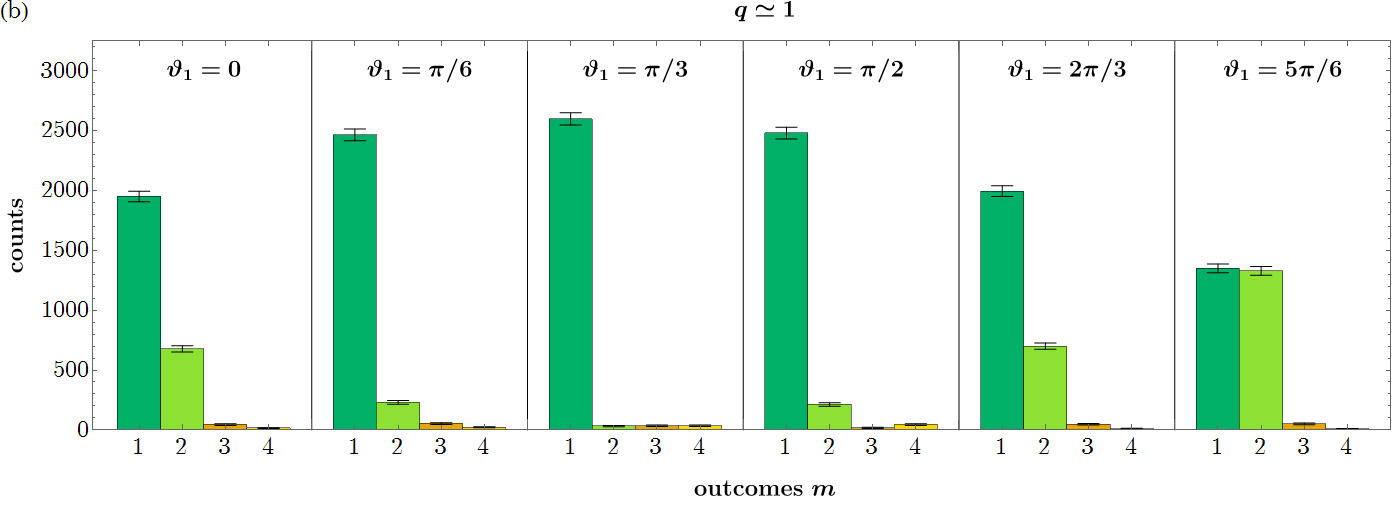}\phantomsubcaption\label{fig:BnoiseProj}
	\end{subfigure}
	\caption{Neutron counts (a) $I_{a,m}$ and (b) $I_{b,m}$ for Pauli observables $A=\vec{a}\cdot\vec{\sigma}$ and $B=\vec{b}\cdot\vec{\sigma}$ where $\vec{a} \simeq \vec{e}_z$, $\vec{b} \simeq \cos(\frac{\pi}{3})\vec{e}_z + \sin(\frac{\pi}{3})\vec{e}_y$, and measurements $\mathcal{M} = \left\{q P_+(\vec{r}_1),q P_-(\vec{r}_1),(1-q) P_+(\vec{r}_2),(1-q)  P_-(\vec{r}_2) \right\}$ with $q\simeq 1$, $\vec{r}_1 = \cos \vartheta_1 \vec{e}_z + \sin \vartheta_1 \vec{e}_y$, and $\vec{r}_2=\vec{e}_z$. This scenario corresponds to that shown in figure~\ref{fig:PlotResults2}(b) of the main text.
	The counts for outcomes $m=3,4$, which ideally should not occur, are due to the fact that the effective value of $q$ is not exactly 1, due primarily to the imperfect purity of the neutron spin states.
	For both (a) and (b), the counts shown are for the positive eigenstate being prepared (i.e., $\ket{+a}$ and $\ket{+b}$, respectively), and the counts are proportional to the joint probabilities $p(a,m)$ and $p(b,m)$. Counts are shown for 6 different measurements, each with $q\simeq 1$ and starting with polar angle $\vartheta_1=0$ and increasing in increments of $\Delta \vartheta_1 = \frac{\pi}{6}$.
	In (a), almost all counts are initially for the $m=1$ outcome, since $\vartheta_1=0$ corresponds to a projective measurement $P_\pm(\vec{r}_1 = \vec{a})$, which ideally would yield this outcome deterministically.
	As $\vartheta_1$ is increased the direction of projection no longer aligns with $\vec{a}$ and outcome $m=2$ becomes more likely. 
	At $\vartheta_1=\frac{\pi}{2}$ the first two outcomes are equiprobable, while for larger $\vartheta_1$ outcome $m=2$ begins to dominate.
	In (b), the measurement is (ideally) deterministic when $\vartheta_1=\frac{\pi}{3}$ since, for this angle $\vec{r}_1=\vec{b}$. The distribution changes symmetrically around the $\vartheta_1=\frac{\pi}{3}$ case, with outcomes $m=1$ and $m=2$ equiprobable when $\vartheta_1=\frac{5\pi}{6}$.
	  }
	\label{fig:DataPVM}
\end{figure}

\begin{figure}[!ht]
	\begin{subfigure}[c]{1.0\textwidth}		
		\includegraphics[width=0.93\textwidth]{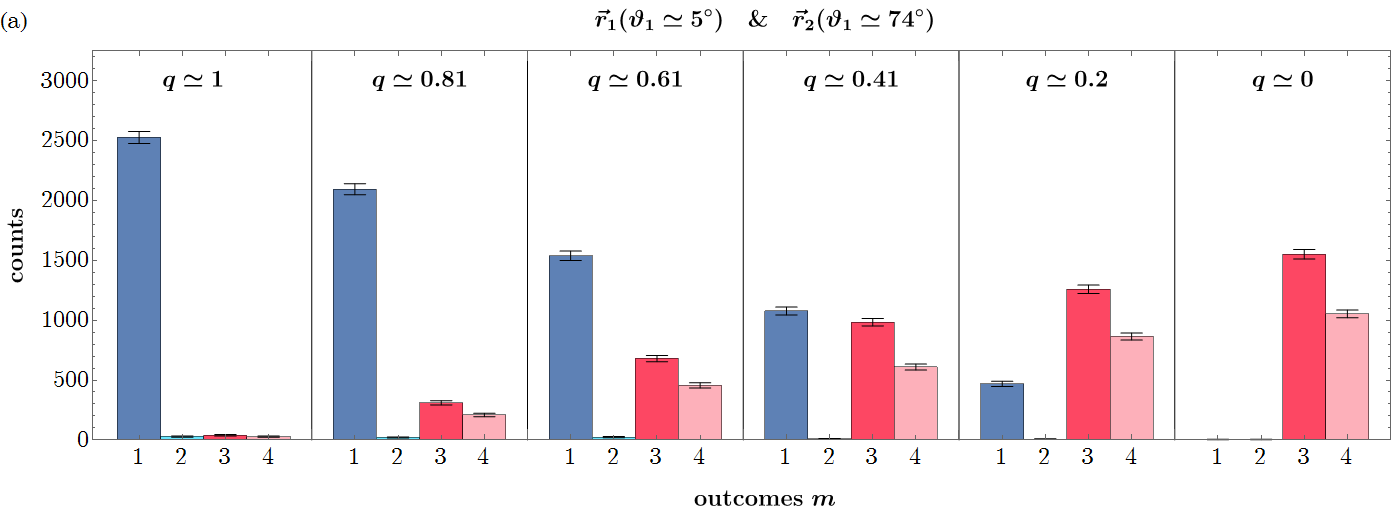}\phantomsubcaption\label{fig:AnoisePOVM}\newline
	\end{subfigure}
	\begin{subfigure}[c]{1.0\textwidth}
		\includegraphics[width=0.93\textwidth]{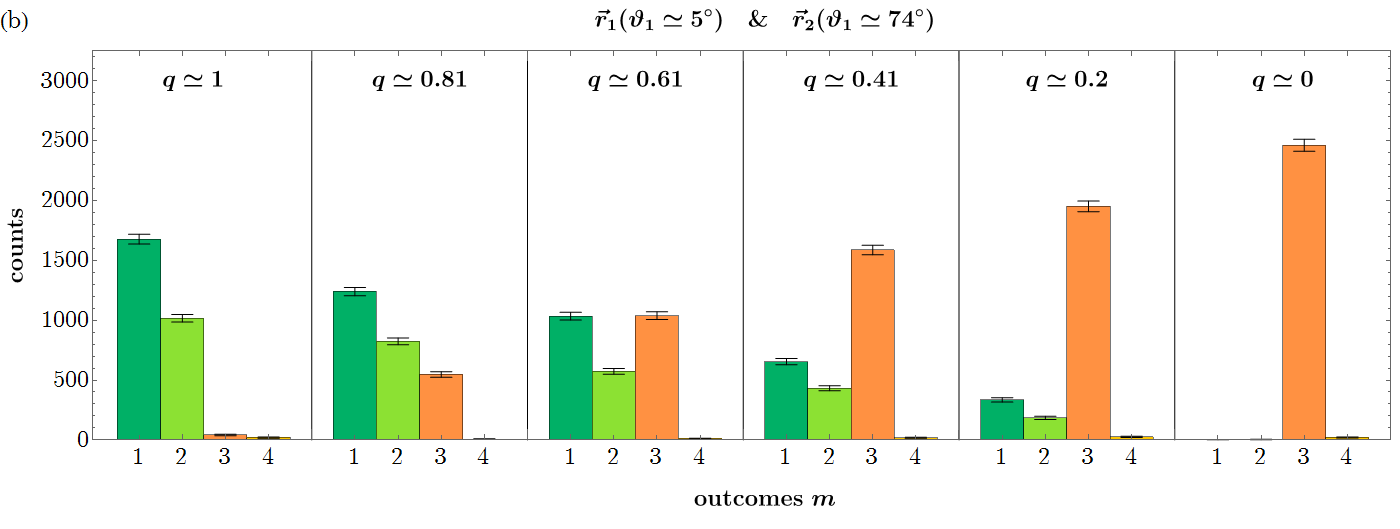}\phantomsubcaption\label{fig:BnoisePOVM}
	\end{subfigure}
	\caption{ Neutron counts (a) $I_{a,m}$ and (b) $I_{b,m}$ for Pauli observables $A=\vec{a}\cdot\vec{\sigma}$ and $B=\vec{b}\cdot\vec{\sigma}$ where $\vec{a} \simeq \vec{e}_z$, $\vec{b} \simeq \cos(79^{\circ})\vec{e}_z + \sin(79^{\circ})\vec{e}_y$, and measurements $\mathcal{M} = \left\{q P_+(\vec{r}_1),q P_-(\vec{r}_1),(1-q) P_+(\vec{r}_2),(1-q)  P_-(\vec{r}_2) \right\}$ with $\vec{r}_1 (\vartheta_1 \simeq 5^{\circ})$, $\vec{r}_2 (\vartheta_2 \simeq 74^{\circ})$ and a range of values of $q$ from $q\simeq 1$ (leftmost box) to $q\simeq 0$ (rightmost box) in decrements of $\Delta q\simeq-0.2$. 
	This scenario corresponds to that shown in figure~\ref{fig:PlotResults}(c) of the main text.
	For both (a) and (b), the counts shown are for the positive eigenstate being prepared (i.e., $\ket{+a}$ and $\ket{+b}$, respectively), and the counts are proportional to the joint probabilities $p(a,m)$ and $p(b,m)$.
	In (a), when $q\simeq 1$ and $q\simeq 0$, the measurements are projective in directions $\vec{r}_1$ and $\vec{r}_2$, respectively. In between the distributions are a convex mixture of these two cases, and thus more than 2 measurement outcomes occur.
	As $q$ is decreased, the measurement transitions from a projective measurement in direction $\vec{r}_1$ -- which is only $5^\circ$ from $\vec{a}$ giving (in theory) a $99.8\,\%$ probability of obtaining outcome $m=1$ -- to a projective measurement in direction $\vec{r}_2$, for which both outcomes $m=3$ and $m=4$ occur. In between, these measurements are mixed and the histogram shows that the joint distribution is a convex mixture of the two extreme cases $q\simeq 1$ and $q\simeq 0$. 
	In (b), the measurement is very close to being deterministic when $q\simeq 0$ since $\vec{r}_2$ is only $5^\circ$ from $\vec{b}$, while for $q\simeq 1$ both outcomes $m=1$ and $m=2$ can occur. As $q$ is varied between these cases, the distribution is again a convex mixture of the two extreme cases, so 3 of the 4 outcomes can occur with non-negligible probability.
	}
	\label{fig:DataPOVM}
\end{figure}

Typical examples of the counts $I_{a,m}$ and $I_{b,m}$ required to determine the entropic noises $N(\mathcal{M},A)$  and $N(\mathcal{M},B)$ are plotted in Figs.~\ref{fig:DataPVM} and~\ref{fig:DataPOVM}. 
The counts shown in Figs.~\ref{fig:AnoiseProj} and~\ref{fig:AnoisePOVM} are proportional to the joint probabilities $p(a,m)$ as well as the conditional probabilities $p(m|a)$ [cf.\ equation~\ref{eq:jointDistFromCnts}].
As a result, when the measurement corresponds to a projective measurement in direction $\vec{r}_1=\vec{a}$ [$\vartheta_1=0$ in figure~\ref{fig:AnoiseProj}], outcome $m=1$ occurs with probability almost 1 (equal in the ideal case). In figure~\ref{fig:AnoiseProj}, as $\vartheta_1$ is increased, the measurement becomes noisy and outcome $m=2$ becomes more probable.
In figure~\ref{fig:AnoisePOVM} when $q\simeq 1$ the measurement is projective along a direction $\vec{r}_1$ very close to $\vec{a}$, while as $q$ is decreased towards 0, $P_\pm(\vec{r}_1)$ is mixed with $P_\pm(\vec{r}_2)$ and outcomes 3 and 4 become more probable as the noisy measurement in direction $\vec{r}_2$ (close to $\vec{b}$) is performed more often by the POVM.
The counts $I_{b,m}$ are interpreted similarly, except that, since the eigenstates of $A$ and $B$ are not aligned, the corresponding distributions are never simultaneously peaked with respect to both observables.

\newpage

\providecommand{\newblock}{}


\begin{thebibliography}{10}
\bibitem{footnote_AM} Note that as we use an entropic definition for the noise, $N(\mathcal{M},A)$ does not depend on the actual values $a$, $m$ taken by $\mathbb{A}$ and $\mathbb{M}$, but only on their distribution.
\end{thebibliography}
\end{document}